\documentclass[aps, prl, reprint, superscriptaddress]{revtex4-1}

\pdfoutput=1 
\usepackage{graphicx}
\usepackage{color,amsmath}
\usepackage[normalem]{ulem}

\usepackage[colorlinks,linkcolor=blue,citecolor=blue,urlcolor=blue]{hyperref}
\usepackage{siunitx}
\usepackage{comment}
\usepackage{layouts}

\begin{document}
\title{A Gate-Defined Quantum Point Contact in an InAs Two-Dimensional Electron Gas}

\author{Christopher Mittag}
\email{mittag@phys.ethz.ch}
\affiliation{Solid State Physics Laboratory, Department of Physics, ETH Zurich, 8093 Zurich, Switzerland}

\author{Matija Karalic}
\affiliation{Solid State Physics Laboratory, Department of Physics, ETH Zurich, 8093 Zurich, Switzerland}

\author{Zijin Lei}
\affiliation{Solid State Physics Laboratory, Department of Physics, ETH Zurich, 8093 Zurich, Switzerland}

\author{Candice Thomas}
\affiliation{Microsoft Station Q Purdue and Department of Physics and Astronomy, Purdue University, West Lafayette, Indiana 47907, USA}
\affiliation{Birck Nanotechnology Center, Purdue University, West Lafayette, Indiana 47907, USA}

\author{Aymeric Tuaz}
\affiliation{Microsoft Station Q Purdue and Department of Physics and Astronomy, Purdue University, West Lafayette, Indiana 47907, USA}
\affiliation{Birck Nanotechnology Center, Purdue University, West Lafayette, Indiana 47907, USA}

\author{Anthony T. Hatke}
\affiliation{Microsoft Station Q Purdue and Department of Physics and Astronomy, Purdue University, West Lafayette, Indiana 47907, USA}
\affiliation{Birck Nanotechnology Center, Purdue University, West Lafayette, Indiana 47907, USA}

\author{Geoffrey C. Gardner}
\affiliation{Microsoft Station Q Purdue and Department of Physics and Astronomy, Purdue University, West Lafayette, Indiana 47907, USA}
\affiliation{Birck Nanotechnology Center, Purdue University, West Lafayette, Indiana 47907, USA}

\author{Michael J. Manfra}
\affiliation{Microsoft Station Q Purdue and Department of Physics and Astronomy, Purdue University, West Lafayette, Indiana 47907, USA}
\affiliation{Birck Nanotechnology Center, Purdue University, West Lafayette, Indiana 47907, USA}

\author{Thomas Ihn}
\affiliation{Solid State Physics Laboratory, Department of Physics, ETH Zurich, 8093 Zurich, Switzerland}

\author{Klaus Ensslin}
\affiliation{Solid State Physics Laboratory, Department of Physics, ETH Zurich, 8093 Zurich, Switzerland}

\date{\today}

\begin{abstract}
We experimentally study quantized conductance in an electrostatically defined constriction in a high-mobility InAs two-dimensional electron gas. A parallel magnetic field lifts the spin degeneracy and allows for the observation of plateaus in integer multiples of  $e^2/h$. Upon the application of a perpendicular magnetic field, spin-resolved magnetoelectric subbands are visible. Through finite bias spectroscopy we measure the subband spacings in both parallel and perpendicular direction of the magnetic field and determine the $g$-factor.
\end{abstract}

\maketitle
When a ballistic two-dimensional charge carrier system is confined to a narrow constriction, a striking quantum mechanical phenomenon can be observed: the conductance is quantized in integer multiples of $2e^2/h$\,\cite{van_wees_quantized_1988,wharam_one-dimensional_1988}. Such a structure, called a quantum point contact (QPC), has since its inception been demonstrated in a plethora of different materials. InAs is a material system featuring a low effective mass and a high spin-orbit interaction, and can be contacted with ease with superconductors due to Fermi level pinning at the surface. This provoked a substantial amount of recent interest in this material for the purposes of topological quantum computation\,\cite{kitaev_unpaired_2001,oreg_helical_2010,lutchyn_majorana_2010,kjaergaard_quantized_2016}. An architecture based on a two-dimensional material system would be beneficial for scaling and integration, which is why full control and understanding of nanostructures in InAs quantum wells has profound implications for future research. However, experiments on QPCs in InAs have been scarce, performed in trench-etched structures\cite{debray_all-electric_2009,martin_field-orientation_2010,lehmann_spin-resolved_2014,matsuo_magnetic_2017}, and were limited by parallel conduction\,\cite{koester_length_1996} due to surface charge carrier accumulation inducing trivial edge conduction\,\cite{noguchi_intrinsic_1991,olsson_charge_1996}. In the following we study a completely electrostatically defined QPC in InAs without any background conductance in order to provide a detailed understanding of its energy levels, $g$-factor and magnetoelectric subband structure.

\begin{figure}[]
	\includegraphics[width=\columnwidth]{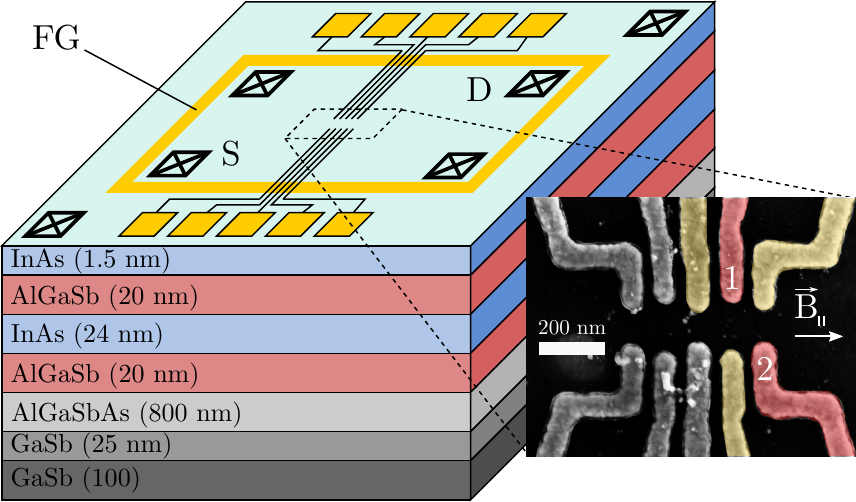}
	\caption{Schematic representation of the lateral sample geometry and heterostructure. The black crossed squares represent Ohmic contacts to the two-dimensional electron gas, the source (S) and drain(D) contacts used in this work are labeled. The frame gate (FG), which is separated from the wafer surface by an $\mathrm{Al}_{\mathrm{2}}\mathrm{O}_{\mathrm{3}}$ layer is represented by the yellow rectangular frame. Another $\mathrm{Al}_{\mathrm{2}}\mathrm{O}_{\mathrm{3}}$ layer separates it from the fine gates forming the nanostructure, which are represented by thin black lines. The inset shows a false-colored scanning electron micrograph of the gate layout of a sample similar to the one used in this work. The two gates forming the QPC studied in the following are colored in red and numbered 1 and 2. The surrounding gates used for investigating the influence of the coupling potential are colored in yellow and are kept on ground except for the measurements in Fig.\,\ref{fig7}.} 
	\label{fig1}
\end{figure} 

The device was fabricated on a heterostructure grown by molecular beam epitaxy on an undoped GaSb substrate in (100) crystal orientation, as described in Ref.\,\onlinecite{thomas_high-mobility_2018}. The schematic cross-section in Fig.\,\ref{fig1} shows the layer sequence, which in growth direction consists of a $\SI{25}{nm}$ GaSb layer, a $\SI{800}{nm}$ Al$_{0.8}$Ga$_{0.2}$Sb$_{0.93}$As$_{0.07}$ quaternary buffer layer, two $\SI{20}{nm}$ Al$_{0.8}$Ga$_{0.2}$Sb barriers below and above the $\SI{24}{nm}$ wide InAs quantum well, and a $\SI{1.5}{nm}$ InAs cap layer. The mobility of the two-dimensional electron gas in the quantum well is $\mu=\SI{1.4e6}{cm^2/Vs}$ at an electron density $n=\SI{5.1e11}{cm^{-2}}$ as measured in a Hall bar geometry at $T=\SI{1.5}{K}$. We determine an elastic mean free path $l_e =\SI{16.5}{\mu\meter}$, which is much larger than the size of the constriction we investigate in the following.

In order to avoid the trivial edge conduction inherent to InAs\,\cite{noguchi_intrinsic_1991,olsson_charge_1996} we refrain from any etching steps and employ a fully gate-defined sample geometry as introduced in Ref.\,\onlinecite{mittag_edgeless_2018} and shown in Fig.\,\ref{fig1}. We begin by depositing Ge/Au/Ni/Au Ohmic contacts on the sample surface, followed by atomic layer deposition (ALD) of a \SI{30}{nm} dielectric layer of $\mathrm{Al}_{\mathrm{2}}\mathrm{O}_{\mathrm{3}}$ at a temperature of \SI{150}{\celsius}. Subsequently, a rectangular (\SI{10/70}{nm}) Ti/Au frame gate is deposited on the surface, followed by another \SI{30}{nm} of $\mathrm{Al}_{\mathrm{2}}\mathrm{O}_{\mathrm{3}}$ grown by ALD. Finally, (\SI{5/25}{nm}) Ti/Au fine gates are deposited on top of the second dielectric layer forming the nanostructure whose shape can be seen in the scanning electron micrograph in the inset of Fig.\,\ref{fig1}. It consists of 5 pairs of opposing gates with separations of \SI{100}{nm} and \SI{150}{nm}. The two gates colored in red are vertically offset by \SI{150}{nm} and horizontally offset by \SI{100}{nm} with respect to the plane of the image and were used to define the QPC investigated in this study. We refer to them as gates QPC1 and QPC2 and they are labelled accordingly in the inset of Fig.\,\ref{fig1}. This configuration showed the best transport data out of all combinations of gates, likely due to a locally favorable potential landscape.  All subsequent measurements were performed in a dilution refrigerator with a base temperature of $T=\SI{60}{mK}$ using low-frequency AC lock-in techniques.

\begin{figure}[]
	\includegraphics{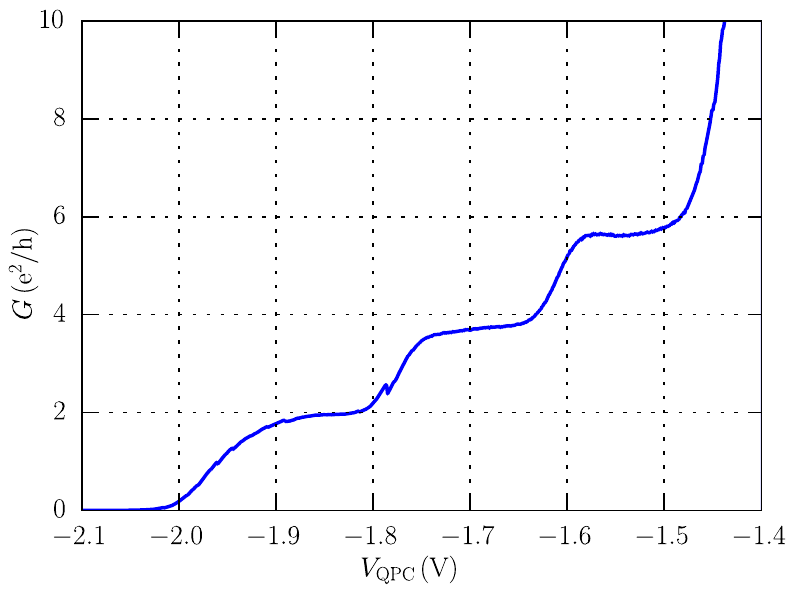}
	\caption{Differential conductance $G$ as a function of $V_{\mathrm{QPC}}$, the voltage applied to the QPC gates, at a temperature of $T=\SI{60}{mK}$ and at zero external magnetic field. Quantized conductance plateaus in steps of $2e^2/h$ are clearly visible.}  
	\label{fig2}
\end{figure}

We bias the frame gate to the voltage $V_{\mathrm{FG}}=\SI{-0.85}{V}$ which is sufficient to deplete the electron gas underneath, thus isolating the inner part of the electron gas from the surrounding electron gas and the sample edges. A constriction can now be defined in the central region using the two fine gates colored in the inset of Fig.\,\ref{fig1} which create the QPC. We apply a fixed AC voltage bias of $\mathrm{d}V_{\mathrm{AC}}= \SI{10}{\micro\volt}$ between the source (S) and drain (D) contact, measure the current $\mathrm{d}I_{\mathrm{AC}}$ and thereby determine the two-terminal differential conductance $G=\mathrm{d}I_{\mathrm{AC}}/\mathrm{d}V_{\mathrm{AC}}$. Fig.\,\ref{fig2} shows $G$ as a function of the voltage applied to the QPC gates $V_{\mathrm{QPC}}$, displaying three plateaus of conductance in steps of $2e^2/h$. The two QPC gates are biased slightly asymmetrically, such that $V_{\mathrm{QPC1}} = V_{\mathrm{QPC}}$ and $V_{\mathrm{QPC2}} = V_{\mathrm{QPC}} + \SI{300}{\milli\volt}$. The visibility of three steps is compatible with an estimate of the number of modes observable by comparing the lithographic width of the QPC to half the Fermi wavelength. The QPC pinches off the electron gas completely and no subtraction of a background conductance was needed, in contrast to previous experiments\,\cite{koester_length_1996,debray_all-electric_2009}. The resistance at pinch-off is at least $10^8\,\Omega$ and the curves showed small hysteresis of the order of $\SI{10}{\milli\volt}$ between up and down sweep direction.
We note that, despite the high quality of the material at hand, we did not observe a feature corresponding to an 0.7 anomaly\,\cite{thomas_possible_1996}.


\begin{figure}[]
	\includegraphics{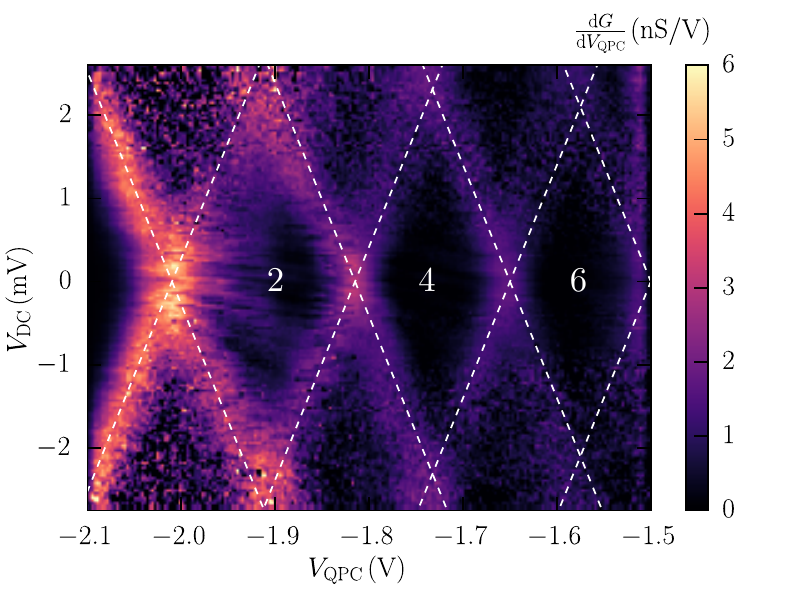}
	\caption{Finite bias spectroscopy showing the transconductance $\mathrm{d}G/\mathrm{d}V_{\mathrm{QPC}}$ as a function of $V_{\mathrm{QPC}}$ and $V_{\mathrm{DC}}$. The data have been corrected for the voltage drop at the series resistance of the contacts. The diamond-shaped stripes of finite transconductance correspond to transitions between plateaus and are highlighted by white dashed lines, delimiting the dark plateau regions which are labeled with the associated conductance in units of $e^2/h$.}  
	\label{fig3}
\end{figure}

In order to map out the energy levels of the QPC, we perform finite bias spectroscopy by applying an additional DC bias voltage between source and drain leading to a bias voltage $V_{\mathrm{DC}}$ across the QPC. Fig.\,\ref{fig3} depicts the numerical derivative with respect to $V_{\mathrm{QPC}}$ of the differential conductance $\mathrm{d}G/\mathrm{d}V_{\mathrm{QPC}}$, as a function of $V_{\mathrm{DC}}$ and $V_{\mathrm{QPC}}$. The data have been corrected to account for the voltage drop across the Ohmic contacts and then rescaled onto the resulting bias axis. For this, an I-V characteristic of the contacts alone without operating any fine gates has been recorded and the voltage drop at the contacts as a function of the total current through the sample has been determined. This voltage is then subtracted from the applied DC bias voltage to determine $V_{\mathrm{DC}}$, the actual bias voltage across the QPC. 

The dark areas correspond to regions where the conductance does not change as a function of $V_{\mathrm{QPC}}$, i.e. conductance plateaus, and the bright regions mark transitions between them. The resulting diamond-shaped structure visualizes the magnitude of the mode spacings in energy, and the slope which has been highlighted by dashed white lines delivers a calibration from $V_{\mathrm{QPC}}$ to energy. The extent of the diamonds in $V_{\mathrm{DC}}$ direction decreases for increasing mode index $n$, corresponding to decreasing energy spacings of the modes. These were determined to be $\Delta E_{12}=\SI{2.73}{\milli\electronvolt}$, $\Delta E_{23}=\SI{2.31}{\milli\electronvolt}$, and $\Delta E_{34}=\SI{2.1}{\milli\electronvolt}$. To assess the validity of these energies, we calculate the real space extent of the modes $L_n$ assuming an harmonic oscillator potential of the frequency $\omega_0$, according to
\begin{equation*}
\tfrac{m^*}{2}\omega_0^2 \cdot L_n^2 = \hbar \omega_0 (n-\tfrac{1}{2}),
\end{equation*}
with the electron effective mass $m^*$. Using $\hbar \omega_0 = \Delta E_{nm}$ for the energy spacings of the respective modes determined from our previous finite bias measurements, we find the length scales $L_1=\SI{32}{\nano\meter}$, $L_2=\SI{53}{\nano\meter}$, and $L_3=\SI{72}{\nano\meter}$. These length scales are within expectation considering the geometric dimensions of the constriction used in this experiment.

\begin{figure}[]
	\includegraphics{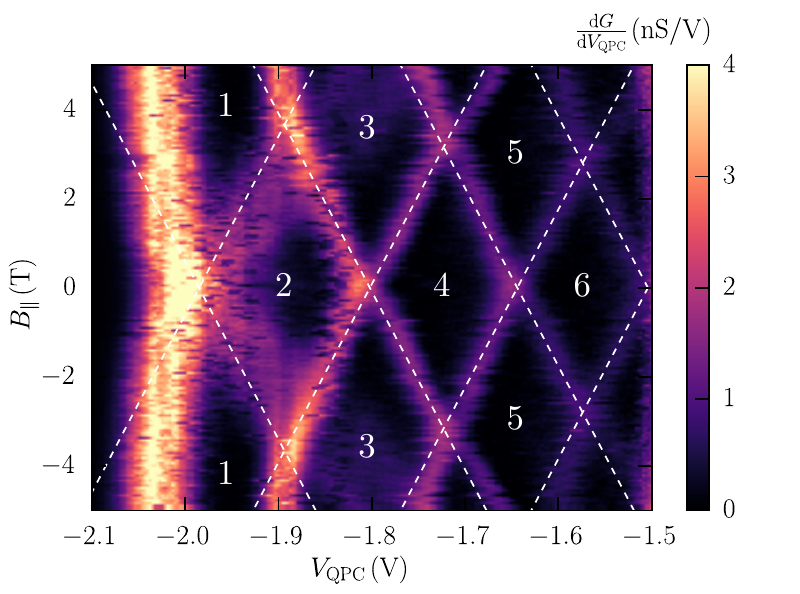}
	\caption{Transconductance $\mathrm{d}G/\mathrm{d}V_{\mathrm{QPC}}$ as a function of $V_{\mathrm{QPC}}$ and the magnetic field $B_{\parallel}$ applied in plane of the two-dimensional electron gas along the direction of the QPC. The spin-degenerate conductance plateaus split according to the Zeeman effect and are labeled with the associated conductance in units of $e^2/h$.}  
	\label{fig4}
\end{figure}

As a next step, we apply a magnetic field $B_{\parallel}$ in parallel to the plane of the two-dimensional electron gas and along the transport direction of the QPC as depicted in the inset of Fig.\,\ref{fig1} and investigate the resulting spin splitting of the modes. Fig.\,\ref{fig4} shows the transconductance $\mathrm{d}G/\mathrm{d}V_{\mathrm{QPC}}$ as a function of $V_{\mathrm{QPC}}$ and $B_{\parallel}$. The three diamond-shaped regions centered along a line through $B_{\parallel}=\SI{0}{\tesla}$ correspond to the spin-degenerate conductance plateaus of the first three modes of the QPC. When increasing $B_{\parallel}$ these modes split and spin-polarized plateaus emerge as additional dark, diamond-shaped regions arranged in a regular pattern around the spin-degenerate plateaus. From the extent of the diamonds in $B_{\parallel}$ and the magnitude of their energy spacings extracted from the finite bias measurements shown in Fig.\,\ref{fig3} we can determine the $g$-factor leading to the Zeeman splitting $\Delta E=g\mu_{\mathrm{B}}\Delta B$. We extract a value of $\vert g \vert=12.6$, closer to the bulk value $g=-15$ of InAs\,\cite{konopka_conduction_1967,pidgeon_interband_1967} than previous measurements on InAs nanowire quantum dots\,\cite{fasth_direct_2007} or etched QPCs\,\cite{matsuo_magnetic_2017}. Using an expression obtained from $\vec{k} \cdot\vec{p}$ theory\,\cite{lassnig_kensuremathrightarrowensuremathcdotpensuremathrightarrow_1985}, we can calculate the effective $g$-factor $g^*$ while taking into account the band-edge parameters of InAs and the quantization energies arising due to the finite quantum well width. This yields a value of  $g^*=-13.6$, which corroborates the experimental results as the quantum well investigated in this study is much wider than the ones used in previous experiments\,\cite{matsuo_magnetic_2017,shabani_gating_2014} and thus shows a larger $g$-factor.

\begin{figure}[]
	\includegraphics{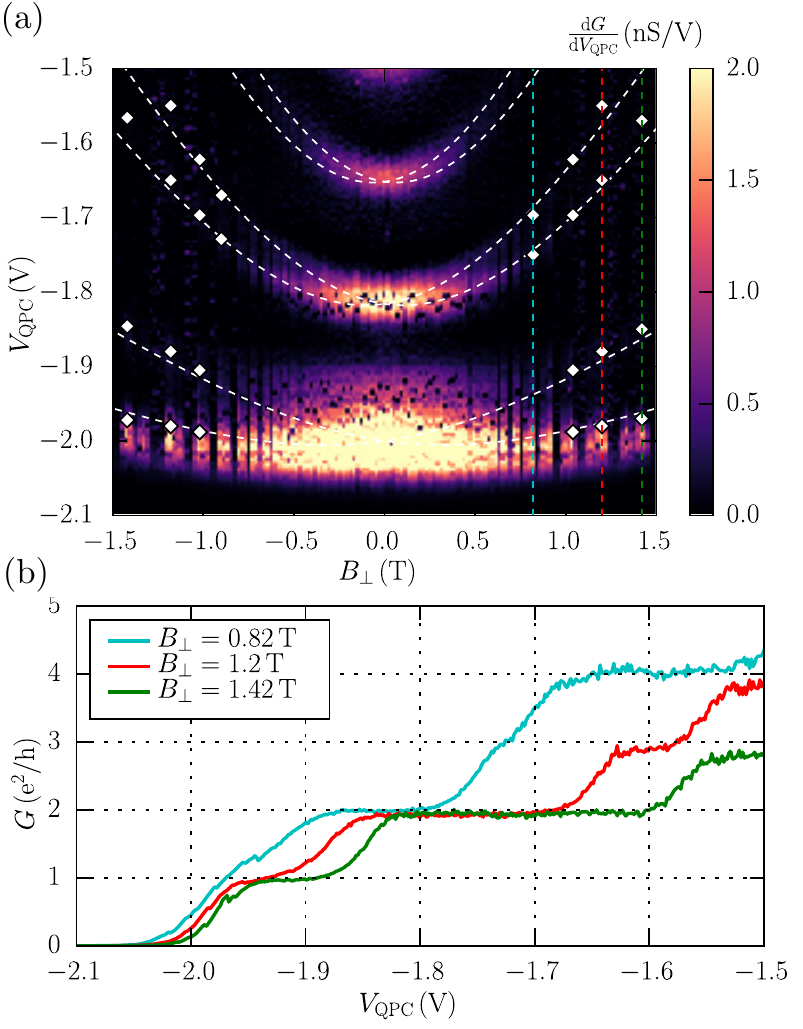}
	\caption{(a) Transconductance $\mathrm{d}G/\mathrm{d}V_{\mathrm{QPC}}$ as a function of $V_{\mathrm{QPC}}$ and $B_{\perp}$. Both magnetic depopulation and spin splitting are visible, and the white dashed lines show a fit to a model incorporating both for the first three modes of the QPC. The dark vertical stripes correspond to regions where the contacts are decoupled due to edge channels. The colored dashed lines correspond to the line cuts shown in (b). The white diamonds mark transitions between plateaus and are extracted from line cuts of the differential conductance. (b) Differential conductance $G$ of the QPC as a function of $V_{\mathrm{QPC}}$ along the dashed lines of equal color shown in (a). Spin split conductance plateaus in integer multiples of $e^2/h$ are emerging.}  
	\label{fig5}
\end{figure}

Now we investigate the effect of a magnetic field perpendicular to the plane of electron gas, $B_{\perp}$, in order to demonstrate the magnetic depopulation of the quantized one-dimensional subbands. In Fig.\,\ref{fig5}\,(a) the transconductance $\mathrm{d}G/\mathrm{d}V_{\mathrm{QPC}}$ of the QPC is depicted as a function of  $V_{\mathrm{QPC}}$ and $B_{\perp}$. The three bright regions along a line through $B_{\perp}=\SI{0}{\tesla}$ mark the transitions between the first three, spin-degenerate modes of the QPC. For increasing $B_{\perp}$ they curve towards higher energy in a parabolic fashion and then transition into a linear slope as they merge with the Landau levels forming in high field. We can observe two parabolas emerging from a single step at $B_{\perp}=\SI{0}{\tesla}$, which are the magnetoelectric subbands\,\cite{van_wees_quantized_1988-1}. A fit of these following a model by Beenakker and van Houten\,\cite{beenakker_quantum_1991} 
\begin{equation*}
E_n= E_0 + (n-\tfrac{1}{2})\cdot \hbar\sqrt{\omega_0^2+\omega_c^2} + g\mu_{\mathrm{B}}B_{\perp},
\end{equation*}
where $\omega_c$ is the cyclotron frequency, and an energy offset $E_0$ is superimposed on the data as white dashed lines. The white diamonds mark transitions between plateaus and were extracted from line cuts of $G$ as a function of $V_{\mathrm{QPC}}$. They serve as a guide to the eye in the regions where $B_{\perp}$ is large and the magnitude of the transconductance is small. In order to better visualize the spin splitting occurring in perpendicular field, we show in Fig.\,\ref{fig5}\,(b) line traces of the differential conductance as a function of $V_{\mathrm{QPC}}$ at three different values of $B_{\perp}$ depicted by the colored dashed lines in Fig.\,\ref{fig5}\,(a). At $B_{\perp}=\SI{0.82}{\tesla}$, the plateau corresponding to the first magnetoelectric subband is not yet visible due to the broad transition to the first mode of the QPC, and the plateau at $G=3e^2/h$ is a narrow shoulder. For $B_{\perp}=\SI{1.2}{\tesla}$, clear steps in units of $e^2/h$ are visible, and for $B_{\perp}=\SI{1.42}{\tesla}$ and higher fields these plateaus become wider and more pronounced.

The vertical black lines in Fig.\,\ref{fig5}\,(a) correspond to regions where no signal was measured due to a finite spacing between the Ohmic contacts and the edge of the frame gate. Therefore, at certain values of $B_{\perp}$ where the bulk enters a fully developed quantum Hall state, the contacts become decoupled from the rest of the sample\,\cite{faist_interior_1991} and the signal vanishes periodically in $1/B_{\perp}$.

\begin{figure}[]
	\includegraphics{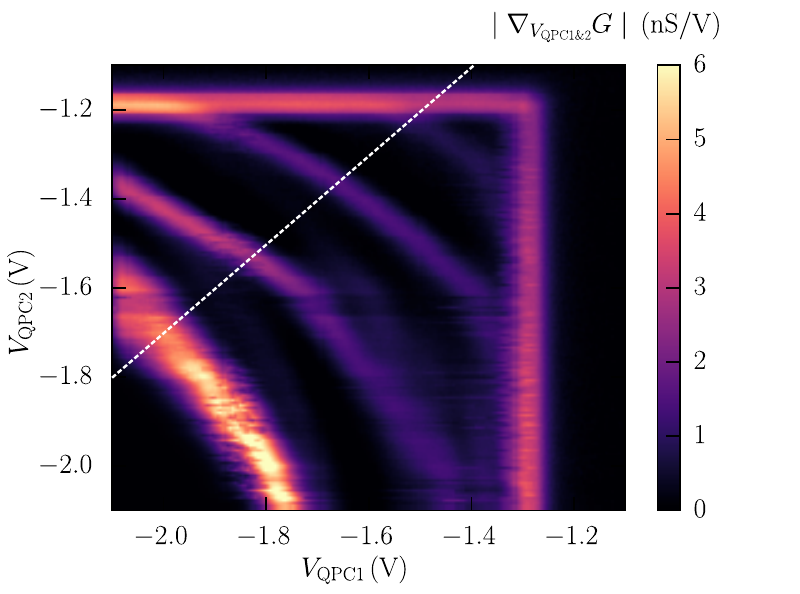}
	\caption{Absolute value of the gradient of the differential conductance $\vert \nabla G \vert$ with respect to the voltages of the two gates forming the QPC, $V_{\mathrm{QPC1}}$ and $V_{\mathrm{QPC2}}$, as a function of these two voltages. Transitions between plateaus appear as arc shaped bright lines. The vertical and horizontal lines correspond to the onset of depletion underneath QPC1 and QPC2 respectively. The white dashed line marks the path in gate voltage space along which all other measurements in this work were recorded.}  
	\label{fig6}
\end{figure}

In order to further characterize the QPC, we laterally shift it by individually changing the voltages applied to the gates forming the constriction. The result can be seen in Fig.\,\ref{fig6} where we plot the absolute value of the gradient of the differential conductance $\vert \nabla G \vert$ with respect to the voltages applied to the two QPC gates, $V_{\mathrm{QPC1}}$ and $V_{\mathrm{QPC2}}$ as a function of these two voltages. Semicircular bright features correspond to transitions between conductance plateaus and show that they are influenced evenly by both gates. The bright horizontal and vertical line, which only depend on $V_{\mathrm{QPC1}}$ or $V_{\mathrm{QPC2}}$ correspond to the pinch-off underneath the respective gate. Impurities in the QPC can be seen as resonances in this type of plot\,\cite{williamson_quantum_1990}, and their slopes correspond to their capacitances to the different QPC gates. Hints of noise are visible on the transition to the first plateau. For all measurements in this paper we chose to sweep the gates along the path in gate voltage space marked by the white dashed line in Fig.\,\ref{fig6}. 
We note that compared to GaAs QPCs with similar scattering times\,\cite{van_wees_quantized_1988,wharam_one-dimensional_1988} the quantization of the plateaus is worse and noise is more pronounced, presumably caused by a background disorder potential. Similar conclusions have been drawn from measurements on quantum Hall states in InAs quantum wells\,\cite{ma_observation_2017}. The reason for this remains an open question and will need to be the subject of future investigation.

\begin{figure}[]
	\includegraphics{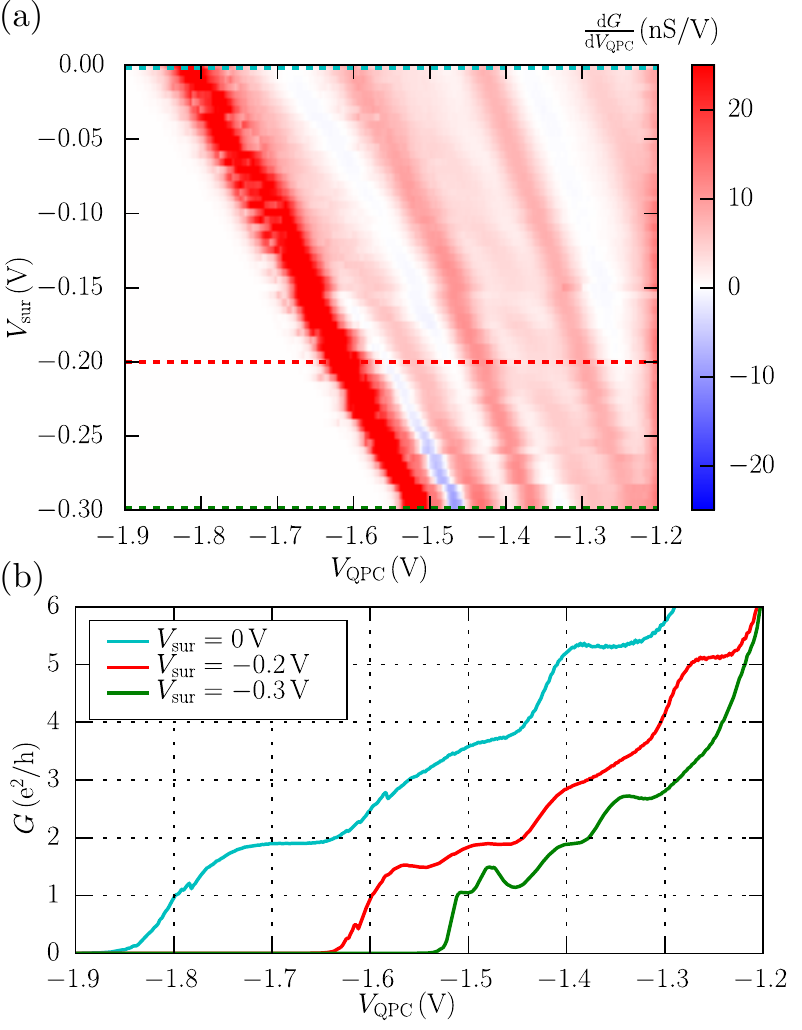}
	\caption{(a) Transconductance $\mathrm{d}G/\mathrm{d}V_{\mathrm{QPC}}$ as a function of $V_{\mathrm{QPC}}$ and $V_{\mathrm{sur}}$ the voltage applied to the fine gates surrounding the QPC which are colored yellow in the inset of Fig.\,\ref{fig1}. As $V_{\mathrm{sur}}$ becomes more negative, additional plateaus (dark regions) emerge. (b) Line cuts of the differential conductance $G$ of the QPC as a function of $V_{\mathrm{QPC}}$ along the dashed colored lines in (a) show the additional plateau-like features which are not quantized to any integer multiple of $e^2/h$.}  
	\label{fig7}
\end{figure}

In etched InAs QPCs, plateaus quantized at odd integer multiples of $e^2/h$ at zero external magnetic field have been reported\,\cite{debray_all-electric_2009,lehmann_spin-resolved_2014,matsuo_magnetic_2017}. We use the tunability of our structure by lateral gates to look for possible origins of such quantization, and to shine some light on the deviations of some of our plateaus from quantization at precise integer multiples of $2e^2/h$. By tuning the voltage on nearby gates surrounding the QPC we change the potential landscape around the QPC and thereby influence the coupling of electron waves from the contact regions into the channel. More specifically, we change the voltage $V_{\mathrm{sur}}$ applied simultaneously to the three neighboring gates of the QPC (colored yellow in the inset of Fig.\,\ref{fig1}) and $V_{\mathrm{QPC}}$ and plot the resulting transconductance of the QPC at zero external magnetic field in Fig.\,\ref{fig7}\,(a). For decreasing $V_{\mathrm{sur}}$, the transitions between plateaus shift to more positive $V_{\mathrm{QPC}}$ due to cross-capacitance, as expected. Furthermore, the onset to the first plateau splits at $V_{\mathrm{sur}}=\SI{-0.15}{\volt}$ and a new conductance plateau arises. Other smooth changes of the conductance as a function of $V_{\mathrm{QPC}}$ are evident in Fig.~\ref{fig7}\,(a).

As examples, line cuts of the conductance at three values of $V_{\mathrm{sur}}$ corresponding to the colored dashed lines in Fig.\,\ref{fig7}\,(a) are displayed in Fig.\,\ref{fig7}\,(b). They show how the conductance traces are continuously deformed as $V_{\mathrm{sur}}$ changes. For example, the red trace ($V_{\mathrm{sur}}=\SI{-0.2}{V}$) shows a new conductance plateau seemingly quantized at $1.6e^2/h$, and the green trace ($V_{\mathrm{sur}}=\SI{-0.3}{V}$) resembles a reentrant plateau feature at $V_{\mathrm{QPC}}=\SI{-1.45}{\volt}$ quantized at about $e^2/h$. Also the precise values of the conductance on plateaus quantized near integer multiples of $2e^2/h$ changes slightly as a function of $V_{\mathrm{sur}}$.

These measurements exemplify how spurious plateau-like features at unconventional conductance values can be generated by suitably molding the potential landscape in the vicinity of the quantum point contact constriction. They show not only that adiabatic coupling of the QPC channel to the leads is of crucial importance, but also that this adiabaticity can be easily spoiled in our InAs samples. In comparison to fully etched constrictions, the degree of tunability provided by the additional gates around the QPC allow us to access these otherwise not attainable regimes. While adiabaticity of the coupling is related to the long-range variations of the potential, resonant structures in the energy-dependent transmission (see green curve in Fig.~\ref{fig7}\,(b)) can be caused by shorter range static disorder\,\cite{mceuen_resonant_1990} in the vicinity of the constriction. Additionally, time-dependent fluctuators may lead to noise and broaden transitions between plateaus as observed for the first plateau in our sample. Our results shown in Fig.~\ref{fig7} may be relevant for the interpretation of previously observed plateaus quantized at odd integer multiples of $e^2/h$ at zero external magnetic field\,\cite{debray_all-electric_2009,lehmann_spin-resolved_2014,matsuo_magnetic_2017}.


In conclusion, we have presented measurements of quantized conductance in a fully gate-defined QPC in an InAs quantum well. In both parallel and perpendicular magnetic field, spin-resolved transport through the constriction could be observed, and the $g$-factor as well as the energy spacing of the modes could be determined. Further optimizing both sample design and growth with the goal of eliminating noise and fluctuations in these nanostructures is needed to pave the way towards reliable operation of more complex nanostructures such as single or double quantum dots. Realizing spin qubits in laterally defined InAs quantum dots will enable performing qubit operations via the spin-orbit interaction similar to nanowires\,\cite{flindt_spin-orbit_2006,nadj-perge_spinorbit_2010} while benefiting from easier integration due to their two dimensional nature.

%
   
\begin{acknowledgments}
The authors acknowledge the support of the ETH FIRST laboratory and the financial support of the Swiss Science Foundation (Schweizerischer Nationalfonds, NCCR QSIT). The work at Purdue was funded by Microsoft Quantum.
\end{acknowledgments}

\bibliography{bibl}
\end{document}